\documentclass[
reprint,
superscriptaddress,
amsmath,
amssymb,
aps,
prl,
]{revtex4-2} 

\usepackage{graphicx}
\usepackage{dcolumn}
\usepackage{bm}
\usepackage[hidelinks]{hyperref}

\usepackage{xcolor}
\usepackage{overpic}
\usepackage{comment}
\usepackage{dsfont}
\usepackage{physics}

\usepackage{cleveref}
\crefname{equation}{Eq.}{Eqs.}
\Crefname{equation}{Equation}{Equations}
\crefname{figure}{Fig.}{Figs.}
\Crefname{figure}{Figure}{Figures}
\crefname{section}{Sec.}{Secs.}
\Crefname{section}{Section}{Sections}
\crefname{appendix}{Appendix}{Appendices}
\Crefname{appendix}{Appendix}{Appendices}

\renewcommand{\d}{\mathrm{d}}
\newcommand{\ee}{\mathrm{e}}
\newcommand{\ii}{\mathrm{i}}

\newcommand{\crefs}[1]{Figs.~\ref{#1}}

\renewcommand{\Tr}[1]{\mathrm{Tr}\left[#1\right]}
\newcommand{\mysec}[1]{\textit{#1---}\!\!}

\begin{document}
\title{Quantum Synchronization of Twin Limit-Cycle Oscillators}

\author{Tobias Kehrer}
\affiliation{Department of Physics, University of Basel, Klingelbergstrasse 82, CH-4056 Basel, Switzerland}
\author{Christoph Bruder}
\affiliation{Department of Physics, University of Basel, Klingelbergstrasse 82, CH-4056 Basel, Switzerland}
\author{Parvinder Solanki}
\affiliation{Institut für Theoretische Physik and Center for Integrated Quantum Science and Technology, Universität Tübingen, Auf der Morgenstelle 14, 72076 Tübingen, Germany}

\date{\today}

\begin{abstract}
Quantum synchronization has been a subject of intensive research in the last decade.
In this work, we propose a quantum Liénard system whose classical equivalent features two limit cycles to one of which the system will converge.
In the quantum case, both limit cycles coexist in a single steady state.
Each of these limit cycles localizes to a distinct phase if coupled to an external drive: one quantum state can thus be assigned two phases.
Furthermore, coupling two such oscillators leads to the simultaneous appearance of synchronization and a synchronization blockade.
To shed light on this apparent paradoxical result, we introduce finer measures of quantum synchronization.
\end{abstract}

\maketitle

\textit{Introduction---}\noindent
Synchronization phenomena of self-sustained oscillating systems have been studied for centuries \cite{Huygens,Buck1938} and remain an active research topic of nonlinear dynamics \cite{Synch_Pikovsky,Synch_Strogatz,Balanov2008,RevModPhys.77.137}.
Self-sustained oscillators exhibit limit cycles, i.e., closed phase-space trajectories that the system converges to regardless of its initial state.
Such limit-cycle oscillators can align quantities like their phase or frequency with other oscillators as well as with external signals.
Quantum analogues of limit-cycle oscillators have been studied in various settings ranging from quantum van der Pol oscillators \cite{PhysRevLett.111.073603,Synch_vdP_Lee,Synch_vdP_Walter,PhysRevE.102.042213,PhysRevResearch.3.013130,PhysRevResearch.5.023021,PhysRevResearch.2.023101} to few-level quantum oscillators \cite{PhysRevLett.121.053601,PhysRevA.101.062104,PhysRevLett.123.023604,cabot2021synchronization}.
While some features of quantum synchronization are inherited from classical synchronization, others originate from genuine quantum properties like entanglement \cite{PhysRevLett.121.063601,PhysRevA.91.012301,PhysRevLett.111.103605,PhysRevE.89.022913,PhysRevA.85.052101,PhysRevLett.129.250601} and (destructive) interference that can lead to a synchronization blockade \cite{PhysRevLett.118.243602,PhysRevA.110.042203,solanki2023blockade}.
Understanding quantum synchronization has implications for quantum sensing \cite{vaidya2024sensing}, quantum thermodynamics \cite{jaseem2020nanoscale,solanki2022role,aifer2024energeticcost}, quantum networks \cite{manzano2013synchronization}, and time crystals \cite{solanki2022seeding,buvca2022algebraic,solanki2024exotic}.
First steps toward its experimental realization have been taken on several platforms including cold atoms \cite{PhysRevLett.125.013601}, nuclear spins \cite{PhysRevA.105.062206}, trapped ions \cite{PhysRevResearch.5.033209}, and superconducting circuits \cite{tao2024noise}.

All studies of quantum synchronization presented up to now consider a single limit cycle.
While classical systems with multiple limit cycles and distinct basins of attraction, known as Liénard systems \cite{Lienard1928,perko2013differential,LienardReview}, have been investigated in detail, synchronization in their quantum analogue has not yet been studied. 
Their amplitude dynamics can be described by an effective potential $V(r)$, see \cref{fig:LC_overview}(a), where the number of limit cycles is given by the number of local minima.
Depending on the initial state, the system converges to one of the limit cycles unless acted on by a noise source that is strong enough to induce switching events.

While there has been an increasing interest in quantum systems featuring multiple separate limit cycles \cite{PhysRevLett.96.103901,Wu2013,Kumar2024,ruby2024lienard,PhysRevE.103.012118,chia2025}, their synchronization properties have not been studied.
In this work, we introduce a Liénard system where two limit cycles \textit{coexist} in a single steady state regardless of the initial conditions and investigate their synchronization behavior.
We call this system a twin limit cycle (TLC):
it is characterized by a double ring-like structure in phase space as sketched in \cref{fig:LC_overview}(b).
The location of the minima (maxima) of the effective potential of the corresponding classical Liénard system is indicated by the dashed (dotted) rings.
Our setup can be extended to host multiple limit cycles.
We examine the synchronization of a single TLC under an external coherent drive and find that the limit cycles exhibit different locking behaviors. 
Furthermore, for two coupled identical TLCs, both synchronization and blockade effects coexist, an apparent paradoxical interplay unattainable with standard limit cycles. 
To distinguish the contributions of individual limit cycles of a TLC, we define new finer measures of quantum synchronization.
Finally, we outline an experimental setup to implement our model.

\begin{figure}[t]
    \centering
    \includegraphics[width=\linewidth]{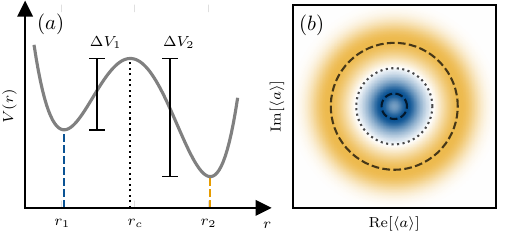}
    \caption{Illustration of a system with multiple limit cycles. (a) Sketch of the effective potential $V(r)$ of a classical Liénard system featuring two basins of attraction separated at $r_c$. (b) Wigner function of a twin limit cycle, i.e., a quantum Liénard system, with $\langle a\rangle = r\ee^{\ii\phi}$. The dashed (dotted) rings with radii $r_1$ and $r_2$ ($r_c$) correspond to the local minima (maximum) of the effective potential obtained from the mean-field equations of motion, see \cref{eq:rMFr}.}
    \label{fig:LC_overview}
\end{figure}

\mysec{Model}
We consider a coherently driven anharmonic quantum oscillator subject to incoherent first and third-order pumping, along with second and fourth-order damping.
These dissipative processes stabilize two concentric limit cycles, see \cref{fig:LC_overview}(b).
The dynamics in the rotating frame of the drive is described by the master equation
\begin{align}
    \dot{\rho} = \mathcal{L}(\rho) =& -\ii[H_0 + H_d, \rho] + \gamma_1\mathcal{D}[a^\dagger]\rho + \gamma_2\mathcal{D}[a^2]\rho \nonumber \\
    &+ \gamma_3\mathcal{D}[a^{\dagger 3}]\rho + \gamma_4\mathcal{D}[a^4]\rho\,,\label{eq:MasterEQ}
\end{align}
where $\mathcal{D}[L]\rho=L\rho L^\dagger - (L^\dagger L \rho + \rho L^\dagger L)/2$ is the Lindblad dissipator, $H_0 = \Delta a^\dagger a^{\phantom{\dagger}} + K a^{\dagger 2} a^{2}$, $H_d = \Omega (a + a^\dagger)$, and $a$ ($a^\dagger$) denote the annihilation (creation) operators of the oscillator.
The detuning between the TLC and the drive is denoted by $\Delta$, $K$ parametrizes the Kerr nonlinearity, and $\Omega$ denotes the strength of the drive.
The rates $\gamma_j$ correspond to incoherent gain (odd $j$) and damping (even $j$).
The model simplifies to the paradigmatic quantum van der Pol oscillator with a single limit cycle for $\gamma_3=\gamma_4=0$.
Additional incoherent processes of higher order, $\mathcal{D}[a^{\dagger n}]\rho$ and $\mathcal{D}[a^{m}]\rho$, with $n(m)$ being odd(even), can be included to obtain multiple limit cycles.

We begin by examining the semiclassical limit to obtain an approximation to the steady state of the quantum system.
The mean-field equations of the effective classical Liénard model can be derived by performing a cumulant expansion to first order.
Setting $\langle a\rangle = r\ee^{\ii \phi}$ results in
\begin{align}
    \dot{r} =& r\left(\frac{\gamma_1}{2} - \gamma_2 r^2 + \frac{3\gamma_3}{2} r^4 - 2\gamma_4 r^6\right) -\Omega\sin(\phi)\,,\label{eq:rMF}\\
    \dot{\phi}=& -\Delta - 2K r^2 -\frac{\Omega}{r}\cos(\phi)\,.\label{eq:phiMF}
\end{align}
Since we are interested in the case of two stable limit cycles, we choose $\gamma_j$ such that the right-hand side of \cref{eq:rMF} exhibits three real zeros $r_1 < r_c < r_2$ at vanishing drive $\Omega=0$,
\begin{align}
    \dot{r} =& r(r_1^2-r^2)(r_c^2-r^2)(r_2^2-r^2)2\gamma_4 \equiv -\partial_r V(r)\,.\label{eq:rMFr}
\end{align}
Here, $r_1$ and $r_2$ are the stable solutions, and $r_c$ is the unstable solution of the mean-field equations, and $V(r)$ is the effective potential, see \cref{fig:LC_overview}(a).
In a classical system, $r_c$ separates the two basins of attraction.
An initial state with $r < r_c$ ($r > r_c$) will therefore converge to $r_1$ ($r_2$).

To explore the corresponding quantum TLC, we examine the Wigner function associated with the steady state of \cref{eq:MasterEQ}. 
The Wigner function exhibits two coexisting concentric limit cycles, as illustrated in \cref{fig:LC_overview}(b), regardless of the initial state. 
In this figure, the dashed and dotted rings represent the stable and unstable solutions of the classical mean-field equation for the oscillator amplitude, respectively, as defined in \cref{eq:rMFr}.
The radius of the outer ring aligns closely with the mean-field prediction $r_2$, while the inner ring shows a notable deviation from $r_1$.

In a classical Liénard system of two limit cycles, both basins of attraction are separated at $r_c$.
However, if extrinsic noise is added, trajectories can cross this boundary.
In the language of the effective potential shown in \cref{fig:LC_overview}(a), noise-induced jumps in $r$ have to overcome the potential barriers $\Delta V_j$ to ``tunnel" between both limit cycles.
In our quantum setup, the steady state is a
combination of two distinct quantum limit cycles.
Considering quantum trajectories, the system tunnels between the two limit cycles due to inherent quantum noise, see Supplemental Material \cite{SuppMat} for a detailed discussion.

\begin{figure}[t]
    \centering
    \includegraphics[width=8.6cm]{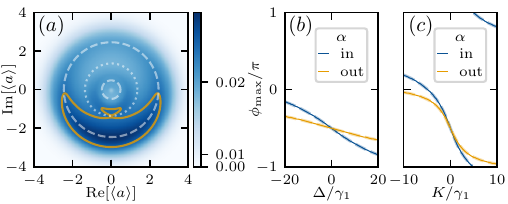}
    \caption{Drive-induced phase locking of a twin limit cycle (TLC).
    (a) Wigner function for $\Omega=8\gamma_1$ and $\Delta=K=0$ showing phase localization of both limit cycles, i.e., a maximum at $\phi = \mathrm{arg}(\langle a\rangle) =\mathrm{arg}(\Omega) - \pi/2 = -\pi/2$.
    Here, dashed (dotted) rings correspond to stable (unstable) solutions of the mean-field equations of the undriven limit cycles, see \cref{eq:rMFr}. 
    The solid orange curves are contour lines at $0.0225$.
    Note the power-law color scale.
    In panels (b) and (c), solid curves denote the maximum of $P_1^\alpha$ with $\alpha\in \{\mathrm{in, out}\}$, and dashed curves denote $\text{arg}(\langle \tilde{a}_\alpha\rangle)$.
    Varying $\Delta$ and $K$, the inner limit cycle exhibits a larger phase shift than the outer one.
    This is opposite to the behavior of a standard quantum van der Pol oscillator, see Supplemental Material \cite{SuppMat}.
    Here, $\Omega=0.25\gamma_1$ for both (b) $K=0$ and (c) $\Delta=0$.
    The dissipation rates for (a)--(c) are $\gamma_2=2.5\gamma_1$, $\gamma_3=1.04\gamma_1$, and $\gamma_4=0.096\gamma_1$.}
    \label{fig:phimax}
\end{figure}

\mysec{One driven twin limit cycle}
We first focus on the synchronization properties of this quantum Liénard system.
First, we discuss phase locking of a TLC to an external drive. 
In \cref{fig:phimax}(a), we present the Wigner function corresponding to the steady state of a driven TLC, which exhibits phase localization near $\phi =\mathrm{arg}(\langle a\rangle) = \mathrm{arg}(\Omega) - \pi/2 = - \pi/2$ indicating the synchronization of both limit cycles to the external drive. 
To characterize and quantify the amount of synchronization, we define a phase localization measure.
Various measures of quantum synchronization have been proposed in the literature
\cite{PhysRevLett.111.073603,PhysRevLett.121.053601,Weiss_2016,phase_dist_Hush,phase_dist_Barak,jaseem2020generalized}.
In this work, we use the synchronization measure based on phase states $\ket{\phi} = \sum_{n=0}^\infty \ee^{\ii n \phi}\ket{n}/\sqrt{2\pi}$  \cite{phase_dist_Barak} where $\ket{n}$ are Fock states, 
\begin{align}
    P_1(\phi) = \bra{\phi}\rho\ket{\phi} - \frac{1}{2\pi}= \frac{1}{2\pi}\sum_{n,m=0}^\infty \ee^{\ii (m-n) \phi} \rho_{nm} - \frac{1}{2\pi}\,,\label{eq:P1def}
\end{align}
with $\rho_{nm}=\bra{n}\rho\ket{m}$.
This measure can be interpreted as a probability distribution of phases $\phi$ from which a uniform distribution is subtracted.
If a state shows no phase preference, this measure will be flat and equal to zero.
For phase-locked oscillators, a single maximum will appear.
Two maxima will be visible for oscillators that exhibit bistable phase locking.
\Cref{eq:P1def} can be further simplified to
\begin{align}
    P_1(\phi) &= \frac{1}{2\pi}\sum_{k=1}^\infty \ee^{-\ii k \phi}\langle\tilde{a}^k\rangle + \mathrm{H.c.}\label{eq:P1viaOp}\,,
\end{align}
where the operator powers $\tilde{a}^k = \sum_{n=0}^\infty \dyad{n}{n+k}$ capture information about the coherence generation and phase localization.
To resolve the phase information of the two limit cycles individually, we define truncated operators $\tilde{a}_\alpha$ with $\alpha\in \{\mathrm{in, out}\}$,
\begin{align}
    \tilde{a}_\text{in} &= \sum_{n=0}^{n_c-1}\dyad{n}{n+1},~~\tilde{a}_\text{out} = \sum_{n=n_c}^{\infty}\dyad{n}{n+1}\,,\label{eq:atilde}
\end{align}
as an approximation of operators that act on the respective subspace of each ring.
The cutoff Fock number $n_c$ is chosen to be the integer closest to $r_c^2$.
We use powers of these $\tilde{a}_\alpha$ to define the phase distributions $P_1^\text{in}$ of the inner and $P_1^\text{out}$ of the outer ring of a TLC,
\begin{align}
    P_1^\alpha(\phi) = \frac{1}{2\pi\langle\mathcal{I}^\alpha\rangle}\sum_{k=1}^\infty \ee^{-\ii k \phi}\langle\tilde{a}_\alpha^k\rangle + \mathrm{H.c.}\, ,\label{eq:P1xviaOp}
\end{align}
where $\alpha \in \{\text{in}, \text{out}\}$.
Here $\mathcal{I}^\alpha$ represents the unit matrix in the subspace $\alpha$,
\begin{align} 
    \mathcal{I}^\text{in} &= \sum_{n=0}^{n_c}\dyad{n}{n},~~\mathcal{I}^\text{out} = \sum_{n=n_c+1}^\infty \dyad{n}{n}\,, \label{eq:IDdef}
\end{align}
and is used to properly normalize the phase distribution.

We use the measure $P_1^\alpha$ to characterize the synchronization properties of the two limit cycles.
In \crefs{fig:phimax}(b) and \ref{fig:phimax}(c), the locking phase angle of a driven TLC is shown for fixed drive strength $\Omega=0.25\gamma_1$, varying detuning $\Delta$, and Kerr nonlinearity $K$.
Remarkably, for nonzero $\Delta$ and $K$, each limit cycle locks to a distinct phase.
Notably, the inner limit cycle responds more strongly to the external drive compared to the outer limit cycle.
This deviates from the mean-field result \cref{eq:phiMF}, which predicts a stronger phase sensitivity of the outer limit cycle, consistent with standard single-limit-cycle oscillators with $\gamma_3=\gamma_4=0$ \cite{SuppMat}.
The higher-order gain and damping channels lead to a distinct dynamical regime that is unique to TLCs.
The overlap of the inner and outer limit cycles of a TLC leads to tunneling and leakage of information between them.
This is a qualitatively different behavior compared with the classical analogue following the mean-field equations.
In addition to the locking of the two limit cycles of a driven TLC to distinct phases shown in \cref{fig:phimax}, we discuss their frequency synchronization in the Supplemental Material \cite{SuppMat}.

\begin{figure*}[t]
	\centering
	\includegraphics[width=\linewidth]{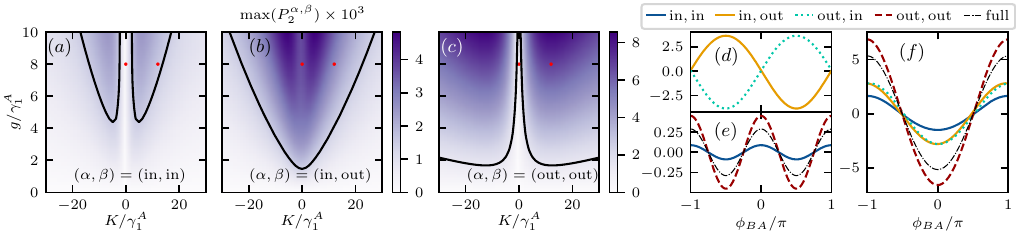}
	\caption{Arnold tongues of two identical coupled twin limit cycles with rates $\gamma^A_2=2.5\gamma^A_1$, $\gamma^A_3=1.04\gamma^A_1$, $\gamma^A_4=0.096\gamma^A_1$, and $\gamma^B_j=\gamma^A_j$ for $\Omega_A=\delta=0$ and $n_c=2$.
		(a)--(c) Maximum of $P_2^{\alpha,\beta}$ as a function of coupling strength and Kerr nonlinearity.
		The black curves denote contour lines at $1.5$.
		(d), (e) Synchronization measures $P_2^{\alpha,\beta}(\phi_{BA})\times 10^3$ corresponding to the left dot in (a)--(c) with $g=8\gamma^A_1$ and $K=0$.
		(f) Similarly, $P_2^{\alpha,\beta}(\phi_{BA})$ for the right dot in (a)--(c) with $g=8\gamma^A_1$ and $K=12/\gamma^A_1$.
		The label `full' corresponds to $P_2(\phi_{BA})$ of \cref{eq:P2viaOp}.
	}
	\label{fig:arnold_tongues_g}
\end{figure*}

\mysec{Two coupled twin limit cycles}
We now focus on the synchronization between two coherently coupled TLCs, described by
\begin{align}
    \dot{\rho} = - \ii [g a^\dagger_A a^{\phantom{\dagger}}_B + \mathrm{H.c.}, \rho] + \mathcal{L}_A(\rho) + \mathcal{L}_B(\rho)\,.\label{eq:MasterEQ2}
\end{align}
Here, the first term is the coherent coupling between the two TLCs with strength $g$.
The Liouvillians $\mathcal{L}_j$ describe the independent dynamics of each TLC similar to \cref{eq:MasterEQ}, where the operators $a_j$ act on oscillator $j$.
In the following, we fix $\Omega_A=\Omega_B=0$, $\delta=\Delta_B - \Delta_A$, and $K=K_A=K_B$.

The phase distribution of two oscillators is obtained by projecting the density matrix onto the tensor products of phase states $\ket{\phi_A, \phi_B}$ and is defined as
\begin{align}
    P_2(\phi_{BA}) &= \int\limits_0^{2\pi}\d\phi\,\bra{\phi,\phi_{BA} + \phi}\rho\ket{\phi,\phi_{BA} + \phi} - \frac{1}{2\pi}\nonumber\\
    &= \frac{1}{2\pi}\sum_{k=1}^\infty \ee^{-\ii k \phi_{BA}}\langle(\tilde{a}^\dagger_A\tilde{a}^{\phantom{\dagger}}_B)^k\rangle + \mathrm{H.c.} \label{eq:P2viaOp}\,.
\end{align}
We integrate over the phase $\phi$ to get the synchronization measure for the relative phase $\phi_{BA}=\phi_{B}-\phi_{A}$.
In analogy to \cref{eq:P1xviaOp}, we define the combined phase distribution $P_2^{\alpha,\beta}$ of two TLCs as
\begin{align}
    P_2^{\alpha,\beta}(\phi_{BA}) =&~ \frac{1}{2\pi\langle\mathcal{I}_A^\alpha\mathcal{I}_B^\beta\rangle}\sum_{k=1}^\infty \ee^{-\ii k \phi_{BA}}\langle\tilde{a}^{\dagger k}_{A,\alpha}\tilde{a}_{B,\beta}^{k\phantom{\dagger}}\rangle+ \mathrm{H.c.}\,,\label{eq:P2xyviaOp}
\end{align}
where $\tilde{a}_{j,\alpha}$ and $\mathcal{I}^\alpha_j$ are the truncated operators $\tilde{a}_{\alpha}$ and the unit operators $\mathcal{I}^\alpha$ that act on the $j$th oscillator.
The measure above allows us to investigate the synchronization between the limit cycles of both oscillators since $(\alpha,\beta)$ can take various combinations: $(\mathrm{in,in})$, $(\mathrm{in,out})$, $(\mathrm{out,in})$, and $(\mathrm{out,out})$.

We now set the dissipation rates $\gamma^A_j=\gamma^B_j$ equal, such that the radii of the inner and outer limit cycles of both TLCs are identical.
In \crefs{fig:arnold_tongues_g}(a) to \labelcref{fig:arnold_tongues_g}(c), we plot the maxima of the combined synchronization measure $P_2^{\alpha,\beta}$. 
For $\delta=0$ and $K=0$, both oscillators have the same frequencies, and hence, we expect a maximum amount of synchronization. 
Surprisingly, the synchronization measure for $(\alpha, \beta)=(\mathrm{in,in})$ and $(\alpha, \beta)=(\mathrm{out,out})$ shown in \crefs{fig:arnold_tongues_g}(a) and \ref{fig:arnold_tongues_g}(c) is highly suppressed around $K=0$.
This is a signature of the synchronization blockade, where the contribution from the first-order locking vanishes ($\langle \tilde{a}^{\dagger}_{A,\alpha}\tilde{a}^{\phantom{\dagger}}_{B,\alpha}\rangle=0$) due to the cancellation of coherences. 
Only second-order phase locking can be observed, as indicated by the two maxima in \cref{fig:arnold_tongues_g}(e).
This blockade can also be understood using the mean-field equations of the relative phase,
\begin{align}
    \dot{\phi}_{BA} =& -\delta + 2K(r_A^2 -r_B^2) + g \frac{r_B^2-r_A^2}{r_A r_B}\cos(\phi_{BA})\,,\label{eq:phiBAMF}
\end{align}
where the coupling term vanishes for limit cycles with equal radii. 
The limit cycles with different radii exhibit an Arnold tongue, see \cref{fig:arnold_tongues_g}(b), signifying synchronization between the $(\mathrm{in,out})$ limit cycles.
Thus, both synchronization and blockade effects occur simultaneously in the coupled identical TLC oscillators, a behavior not known in classical analogues in the absence of noise. 

The synchronization blockade is lifted for $K\neq 0$ \cite{PhysRevLett.118.243602}, as shown in \crefs{fig:arnold_tongues_g}(a) and \ref{fig:arnold_tongues_g}(c).
For $\delta=0$ and $K>0$ ($K<0$) the relative phase of both oscillators locks to $\phi_{BA}=\pi$ ($\phi_{BA}=0$), see \cref{fig:arnold_tongues_g}(f).
To understand this behavior, we investigate the mean-field equations (see Supplemental Material \cite{SuppMat} for more details).
We examine the phase-locking behavior by expanding \cref{eq:phiBAMF} about the radii $r_1$ and $r_2$ of the limit cycles in the mean-field approximation. 
There are two cases: (i) equal radii $r_j = r_x + r_{j,1} g/\gamma_1$ where $j\in\{A,B\}$, $x,y\in\{1,2\}$, and $x\neq y$,
\begin{align}
    \dot{\phi}_{BA} =& -\delta + \frac{4gKr_x^2\sin(\phi_{BA})-g^2\sin(2\phi_{BA})}{2r_x^2(r_x^2-r_y^2)(r_x^2-r_c^2)\gamma_4}\,,\label{eq:phiMFTLCxx}
\end{align}
and (ii) different radii $r_A = r_x + r_{A,1} g/\gamma_1$ and $r_B = r_y + r_{B,1} g/\gamma_1$
\begin{align}
    \dot{\phi}_{BA} =& -\delta + 2K(r_x^2 - r_y^2) + g \frac{r_y^2 - r_x^2}{r_1 r_2}\cos(\phi_{BA})\nonumber\\
    &+ gK\frac{r_1^2+r_2^2-r_c^2}{r_1 r_2(r_c^2-r_1^2)(r_2^2-r_c^2)\gamma_4}\sin(\phi_{BA})\,.\label{eq:phiMFTLCxy}
\end{align}
For $\delta=K=0$, these mean-field equations lead to (i) bistable locking at $\phi_{BA}=0,\pi$ and (ii) locking to a single phase value, like in the full quantum system shown in \crefs{fig:arnold_tongues_g}(d) and \ref{fig:arnold_tongues_g}(e).
In contrast to the quantum case, the relative phase in \cref{eq:phiMFTLCxy} locks to $\phi_{BA}=\pi/2$ for $(\alpha,\beta)=(\mathrm{in},\mathrm{out})$ and to $\phi_{BA}=-\pi/2$ for $(\alpha,\beta)=(\mathrm{out},\mathrm{in})$.
For $\delta=0$ and $K>0$ ($K<0$), the equation of motion \cref{eq:phiMFTLCxx} of the relative phase of equal limit cycles has a single stable solution at $\phi_{BA}=\pi$ ($\phi_{BA}=0$), see \cref{fig:arnold_tongues_g}(f).
For different limit cycles, however, $K>0$ ($K<0$) in \cref{eq:phiMFTLCxy} leads to a shift of the locking phase toward $\phi_{BA}=0$ ($\phi_{BA}=\pi$).
The mean-field analysis is suitable to predict the locking of the relative phases of (i) equal limit cycles of two coupled TLCs but fails to describe the locking of (ii) different limit cycles.
An explanation might be that in the quantum Liénard system, both limit cycles are not strictly separated like the basins of attraction in the classical analogue. 
Therefore, locking mechanisms of different pairs of limit cycles of two TLC oscillators interplay.

For our choice of parameters $n_c=2$ and $\tilde{a}_\mathrm{out} \approx \tilde{a}$, $P_2^{\mathrm{out},\mathrm{out}}$ shown in \cref{fig:arnold_tongues_g}(c) behaves qualitatively similar to the standard synchronization measure $P_2$ defined in \cref{eq:P2viaOp}.
The existence of the blockade at $K=0$ is therefore also confirmed by $P_2$, see \cref{fig:arnold_tongues_g}(e).
In the Supplemental Material \cite{SuppMat}, we furthermore explore the mutual information as a measure of synchronization \cite{PhysRevA.91.012301} and find its behavior to be qualitatively similar to our choice of synchronization measure $P_2^{\alpha,\beta}$.
However, the blockade at $K=0$ is not well captured.

\mysec{Experimental realization}
The effects discussed here can be potentially observed in a trapped-ion experiment similar to \cite{PhysRevLett.131.043605} that demonstrated quantum synchronization of a phonon laser to an external signal.
The setup consists of a calcium and a beryllium ion in a radio-frequency trap that share a common harmonic mode of motion which is denoted by the annihilation operator $a$.
To realize $n$th-order gain ($m$th-order damping), one has to implement a sideband heating (cooling) laser that is detuned from a particular transition in the ion energy-level scheme \cite{RevModPhys.75.281}.
If this detuning equals $n$ ($-m$) times the energy of the harmonic mode and assuming fast ion decay with respect to the timescales of the motion in the trap, an effective jump operator $L=a^{\dagger n}$ ($L=a^m$) is realized.
For each dissipator in \cref{eq:MasterEQ}, a distinct ion transition has to be chosen.
Therefore, to realize the four gain and damping channels of a TLC, two spin transitions per ion have to be driven with one of the four red and blue sideband lasers each.

\mysec{Conclusion}
We have presented a quantum Liénard system whose steady state hosts two coexisting
limit cycles. 
This is qualitatively different in the classical analogue without noise, where the phase space of the oscillator splits in distinct basins of attraction for each limit cycle.
Due to the coexistence of both limit cycles, the quantum system exhibits surprising synchronization behavior: coherently driving this quantum twin limit cycle (TLC) oscillator, each of the two limit cycles locks to a distinct phase when the oscillator is detuned from the drive or in the presence of a Kerr nonlinearity.
Varying the detuning or the Kerr nonlinearity, the inner limit cycle exhibits a larger phase shift than the outer one.
In contrast, the induced phase shift of a standard quantum van der Pol oscillator increases monotonically with its radius.
A pair of coherently coupled identical TLC oscillators shows an apparent paradoxical effect: the relative phase of two \textit{equal-sized} limit cycles of oscillators $A$ and $B$ exhibits bistable locking, i.e., the oscillators are in the quantum synchronization blockade.
Simultaneously, two limit cycles of \textit{different} radius lock to a single value of the relative phase.
Therefore, in a pair of TLCs, both synchronization and blockade \textit{coexist} within the same steady state.
Thus, TLCs exhibit synchronization properties that differ in a qualitative way from those of conventional limit cycle oscillators.
They provide a foundation for exploring complex collective dynamics and enable the understanding of quantum synchronization in more general systems with multiple coexisting attractors.

Our setup can be extended by incorporating higher-order gain and damping channels, leading to multiple local minima in the effective potential and, therefore, multiple limit cycles.
Another choice of dissipation channels that are more localized in Fock space has been studied in \cite{Rips2012}.
Employing such channels will lead to multiple effective few-level limit cycles in Fock space centered at various Fock numbers.
Future directions also include the study of minimal examples, e.g., spin-2 oscillators where both $\ket{\pm1}$ are stabilized, as well as networks of TLCs.
The analysis of multiple limit cycles within a single steady state opens a promising avenue within the field of quantum synchronization with potential applications in quantum sensing and entanglement generation.

\begin{acknowledgments}
    \mysec{Acknowledgments}
    We thank Albert Cabot and Tobias Nadolny for fruitful discussions.
    T.K. and C.B.~acknowledge financial support from the Swiss National Science Foundation individual grant (Grant No.~200020 200481). 
    P.S. acknowledges support from the Alexander von Humboldt Foundation through a Humboldt research fellowship for postdoctoral researchers.
    We furthermore acknowledge the use of \textsc{QuTip} \cite{qutip5} and \textsc{multiprocess} \cite{multiprocess}.
    The data that support the findings of this article are openly available \cite{data}.
\end{acknowledgments}

%

\pagebreak
\clearpage
\onecolumngrid
\onecolumngrid
\begin{center}
	\textbf{\large Supplemental material: Quantum Synchronization of Twin Limit-Cycle Oscillators}
\end{center}
\maketitle

\newcounter{sfigure}
\stepcounter{sfigure}
\renewcommand{\thefigure}{S\arabic{sfigure}}
\newcommand{\scaption}[1]{\stepcounter{sfigure}\caption{#1}}

\setcounter{table}{0}
\setcounter{page}{1}
\newcounter{ssection}
\newcounter{ssubsection}
\makeatletter
\newcommand{\ssection}[1]{\stepcounter{ssection}\section{\arabic{ssection}.~#1}\setcounter{ssubsection}{0}}
\newcommand{\ssubsection}[1]{\stepcounter{ssubsection}\subsection{\arabic{ssection}.\arabic{ssubsection}.~#1}}

\setcounter{equation}{0}
\renewcommand{\theequation}{S\arabic{equation}}

Here, we present some more detailed calculations and derivations, including the analysis of a single quantum trajectory of a twin limit cycle (TLC), the frequency synchronization of a TLC, the predictions of the locking behavior of TLCs based on their mean-field equations, the phase locking of standard quantum limit cycles, regions in TLC parameter space where the synchronization blockade exists, and the quantum mutual information of coupled TLCs.

\begin{figure}[h!]
	\centering
	\includegraphics[width=\linewidth]{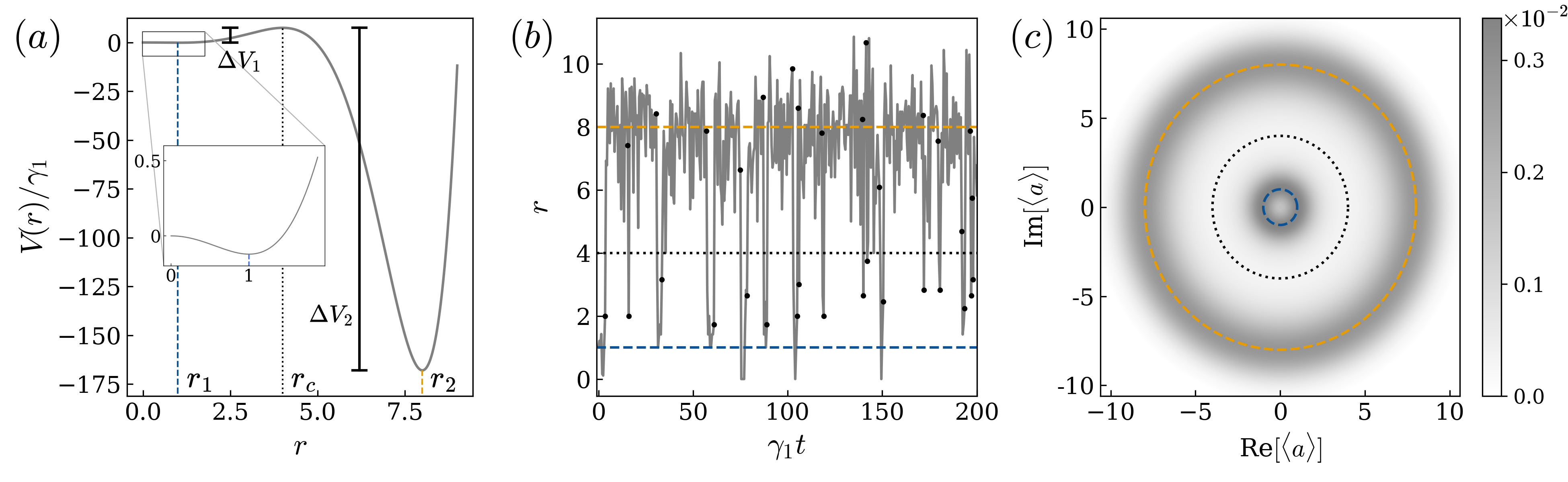}
	\caption{(a) Effective potential of the mean-field equations (see \cref{eq:rMFr} of main text) along the radial direction exhibits minima at $r_1=1$ and $r_2=8$, and a maximum at $r_c=4$.
		Here $\Delta V_{1(2)}$ is the potential difference between the stable minima$r_{1(2)}$ and the unstable maximum $r_c$.
		The inset shows the minimum of $V(r)$ at $r_1$.
		(b) A single trajectory showing the tunneling between two limit cycles with stable radii at $r_1=1$ (blue dashed line) and $r_2=8$ (orange dashed line), separated by the unstable radius at $r_c=4$ (black dotted line).
		The transition point between two limit cycles during the time evolution is indicated by black dots.
		(c) The Wigner function corresponding to the steady state shows the long-time average behavior of many quantum trajectories, resulting in the occurrence of two limit cycles. 
		The rates for all the panels are $\gamma_{2}=0.539\gamma_{1},\gamma_{3}\approx 0.0264\gamma_{1},\gamma_{4}\approx 0.000244\gamma_{1}$ and the cutoff in the Fock basis is $N=120$.}
	\label{fig:trajectory}
\end{figure}

\ssection{Twin limit cycle and quantum trajectory analysis}
In this section, we examine the coexistence of limit cycles by analyzing the dynamics of a single quantum trajectory in a quantum Liénard system. 
In \cref{fig:trajectory}, we specifically choose dissipation rates $\gamma_i$ that allow the effective potential $V(r)$ of the system to have two well-separated stable minima at $r_1 = 1$ (inner limit cycle) and $r_2 = 8$ (outer limit cycle), as well as an unstable maximum at $r_c = 4$.
The minima of $V(r)$ are sufficiently separated to facilitate the observation of the transition from one stable radius to the other due to intrinsic quantum noise.
As shown in \cref{fig:trajectory}(a), the noise has to overcome a smaller potential difference, $\Delta V_1$, when transitioning from $r_1$ to $r_2$, in contrast to the larger potential difference, $\Delta V_2$, for the reverse direction. 
This asymmetry in the potential differences is evident in the time evolution of a single trajectory presented in \cref{fig:trajectory}(b), where the system spends more time in the outer limit cycle at $r_2$ compared to the inner limit cycle at $r_1$.
Notably, this behavior of the quantum trajectories is independent of the choice of the initial state.
The steady-state density matrix can be understood as the long-time average of many quantum trajectories, which results in the two ring-like structures found in the corresponding Wigner function, as shown in \cref{fig:trajectory}(c).

\ssection{Frequency Synchronization}

In this section, we discuss the frequency synchronization of TLCs and compare it with the phase synchronization measure. 
We utilize the power spectrum to analyze the frequency entrainment of the TLC, which is defined as follows
\begin{align}
S_\alpha(\omega) = \lim_{t\to\infty}\int\limits_{-\infty}^\infty \d\tau \langle a^\dagger_\alpha(t+\tau) a_\alpha(t) \rangle \ee^{-\ii \omega\tau}\,, \label{eq:spectrum}
\end{align}
where $\alpha=\text{in},\text{out}$.
The power spectrum defined above is based on two-time correlations of the truncated annihilation operators, which we define as
\begin{align}
a_\text{in} = \sum_{n=0}^{n_c-1}\sqrt{n+1}\dyad{n}{n+1},~~a_\text{out} = \sum_{n=n_c}^{\infty}\sqrt{n+1}\dyad{n}{n+1}\,.\label{eq:truncop}
\end{align}
Note the difference to \cref{eq:atilde}, where the factor $\sqrt{n+1}$ is not included.
\begin{figure}[t]
	\centering
	\includegraphics[width=17.2cm]{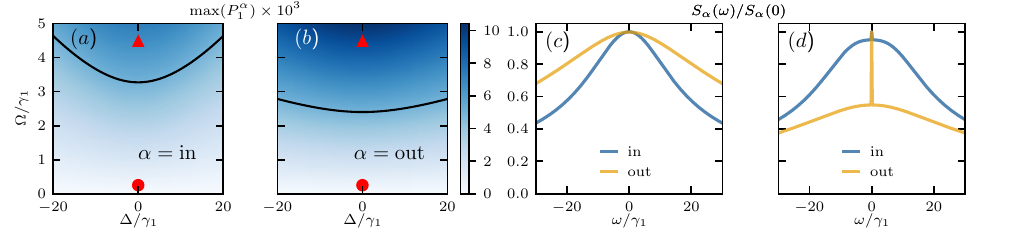}
	\scaption{Phase and frequency synchronization of a driven TLC with $\gamma_2=2.5\gamma_1$, $\gamma_3=1.04\gamma_1$, and $\gamma_4=0.096\gamma_1$.
		(a), (b) Arnold tongues for inner and outer limit cycles as a function of detuning $\Delta$ and drive strength $\Omega$ describing the magnitude of phase synchronization.
		The black curves denote contour lines at $5$.
		(c) Power spectra $S_\alpha(\omega)$ corresponding to the red dot at $\Delta=0$ and $\Omega=0.25\gamma_1$ in panels (a) and (b).
		(d) Power spectra $S_\alpha(\omega)$ corresponding to the red triangle at $\Delta=0$ and $\Omega=4.5\gamma_1$ in panels (a) and (b).
		Both spectra exhibit a peak at $\omega=0$ due to injection locking.
		The power spectra are based on the truncated annihilation operators defined in \cref{eq:truncop}.
	}
	\label{fig:frequency_synch}
\end{figure}
We start by discussing the phase synchronization measure, i.e., the maximum of $P_1^\alpha(\phi)$ defined in \cref{eq:P1xviaOp}. 
The phase synchronization measure shows Arnold tongues for both the inner and outer limit cycle as functions of detuning $\Delta$ and drive strength $\Omega$, as illustrated in \crefs{fig:frequency_synch}(a) and \ref{fig:frequency_synch}(b).
Both limit cycles exhibit a vanishing amount of synchronization for lower $\Omega$ values and require higher coupling strength to observe a large amount of synchronization due to the presence of higher-order dissipators.

The power spectrum is well-defined for both the limit cycles for small values of $\Omega$.
One such example is shown in \cref{fig:frequency_synch}(c) where $\Omega=0.25\gamma_1$ and $\Delta = 0$.
However, no frequency locking occurs when $\Delta \neq 0$ for such smaller drive strength.
In regions of stronger drive, where a significant degree of phase synchronization is observed, the spectrum of the inner limit cycles broadens considerably, as illustrated in \cref{fig:frequency_synch}(d) for $\Omega=4.5\gamma_1$ and $\Delta=0$.
Such spectral broadening results from changes in the population distribution within the inner limit cycle, shifting toward the critical radius for higher $\Omega$ values.
With a further increase in $\Omega$, the population increasingly overlaps with the outer limit cycle, expected to lead to further broadening of the power spectrum. 
This shift hampers the quantization of synchronization within the inner limit cycle when analyzed through the power spectrum, limiting its applicability to a 
narrow region that holds little significance.
Therefore, the phase synchronization measure is more sensitive and applicable even at higher drive strengths for the TLC oscillators.


\ssection{Expansion of the mean-field equations}
In this section, we extend our mean-field analysis to understand the phase-locking behavior of coupled TLC oscillators. 
The mean-field equations for two TLCs are given as follow
\begin{align}
\dot{r}_j =& r_j\left(\frac{\gamma^j_1}{2} - \gamma^j_2 r_j^2 + \frac{3\gamma^j_3}{2} r_j^4 - 2\gamma^j_4 r_j^6\right)+ g r_i \sin(\phi_i-\phi_j)  -\Omega_j\sin(\phi_j)\,,\\
\dot{\phi}_j=& -\Delta_j-2K_j r_j^2 -g\frac{r_i}{r_j}\cos(\phi_i-\phi_j) -\frac{\Omega_j}{r_j}\cos(\phi_j)\,.
\end{align}
The above equations are derived from the Lindblad master equation defined in \cref{eq:MasterEQ2} by setting $\langle a_j\rangle = r_j\ee^{\ii\phi_j}$.
In the following, we expand these equations for identical TLCs with $\gamma^A_j = \gamma^B_j$, $\delta=\Delta_B-\Delta_A$, $K=K_A=K_B$, and $\Omega_A = \Omega_B = 0$ up to second order in $\epsilon = g/\gamma_1$ about the stable radii.

\ssubsection{Limit cycles of the same type: (in,in) and (out,out)}
Here we consider the case of limit cycles of the same type, (in,in) and (out,out), for two TLCs. 
Using \cref{eq:rMFr}, the steady-state solutions close to the stable radii $r_x$ are given by
\begin{align}
r_j = r_x + \epsilon\,r_{j,1} + \epsilon^2\,r_{j,2},~~j\in\lbrace A,B\rbrace,~~x, y\in\lbrace 1,2\rbrace,~~x\neq y\,,
\end{align}
where $\epsilon = g/\gamma_1$ and
\begin{align}
r_{A,1} &= \frac{\sin(\phi_{BA})}{4r_x(r_x^2-r_y^2)(r_x^2-r_c^2)} \frac{\gamma_1}{\gamma_4}= -r_{B,1}\,,\\
r_{A,2} &= \sin[2](\phi_{BA})\frac{9r_x^2(r_y^2+r_c^2)-5r_y^2r_c^2-13r_x^4}{32r_x^3(r_x^2-r_y^2)^3(r_x^2-r_c^2)^3}\frac{\gamma_1^2}{\gamma_4^2} = r_{B,2}\,.
\end{align}
Note that since $r_1 < r_c < r_2$, the product $(r_x^2-r_y^2)(r_x^2-r_c^2)>0$ in the denominators is always positive.
The resulting equation of motion of the relative phase $\phi_{BA}$ when expanding both twin limit cycles about $r_x$, cf.~\crefs{fig:arnold_tongues_g}(a) and \ref{fig:arnold_tongues_g}(c), reads
\begin{align}
\dot{\phi}_{BA} = -\delta + \frac{4gKr_x^2\sin(\phi_{BA})-g^2\sin(2\phi_{BA})}{2r_x^2(r_x^2-r_y^2)(r_x^2-r_c^2)\gamma_4}\,.\label{eq:phiMFTLCxxSupp}
\end{align}
For $\delta=0$, the relative phase between both inner or outer limit cycles locks to a single value $\phi_{BA}=\pi$ ($\phi_{BA}=0$) if $K>0$ ($K<0$).
If $K=0$, bistable locking to $\phi_{BA}=0,\pi$ occurs, which corresponds to the synchronization blockade due to the absence of first-order phase locking.
Thus, the system is in the synchronization blockade between the limit cycles of the same type of the TLCs at $K=0$.
For $K\neq 0$ the blockade is lifted.
This explains the behavior shown in \crefs{fig:arnold_tongues_g}(a,c,e,f) of the main text.

\ssubsection{Limit cycles of different type: (in,out) and (out,in)}
When expanding both TLCs about different stable radii $r_x \neq r_y$, cf.~\crefs{fig:arnold_tongues_g}(b) and \ref{fig:arnold_tongues_g}(d), the steady-state solutions to first order in $\epsilon=g/\gamma_1$ are given by
\begin{align}
r_A &= r_x + \epsilon\,r_{A,1}\,,\\
r_B &= r_y + \epsilon\,r_{B,1}\,,
\end{align}
where
\begin{align}
r_{A,1} &= \frac{r_y\sin(\phi_{BA})}{4r_x^2(r_x^2-r_y^2)(r_x^2-r_c^2)}\frac{\gamma_1}{\gamma_4}\,,\\
r_{B,1} &= \frac{r_x\sin(\phi_{BA})}{4r_y^2(r_x^2-r_y^2)(r_y^2-r_c^2)}\frac{\gamma_1}{\gamma_4}\,.
\end{align}
The equation of motion for the relative phase reads
\begin{align}
\dot{\phi}_{BA} =& -\delta + 2K(r_x^2 - r_y^2) + g \frac{r_y^2 - r_x^2}{r_1 r_2}\cos(\phi_{BA}) + gK\frac{r_1^2+r_2^2-r_c^2}{r_1 r_2(r_c^2-r_1^2)(r_2^2-r_c^2)\gamma_4}\sin(\phi_{BA})\,.\label{eq:phiMFTLCxySupp}
\end{align}
For $\delta=K=0$, the relative phase between limit cycles of different radii, such as (in, out) or (out, in), locks to a single value $\phi_{BA}=\pi/2$ ($\phi_{BA}=-\pi/2$) if $x=1,y=2$ ($x=2,y=1$).
The different limit cycles of the coupled TLCs synchronize, while limit cycles of the same type exhibit the synchronization blockade.
For $K\neq 0$, the coefficient in front of the $\sin(\phi)$ term in \cref{eq:phiMFTLCxySupp} has the sign of $K$.
Together with the second term of \cref{eq:phiMFTLCxySupp}, for $K>0$ ($K<0$) we expect a shift of the maximum towards $\phi=0$ ($\phi=\pm\pi$).
This prediction deviates from the observed locking behavior presented in \crefs{fig:arnold_tongues_g}(d) and \ref{fig:arnold_tongues_g}(f).
An explanation might be that in the quantum Liénard system, the rings in the Wigner function corresponding to the quantum limit cycles overlap, unlike the basins of attraction in the classical analogue that are strictly separated.
Therefore, locking mechanisms of various pairs of limit cycles of two TLC oscillators interplay.

\begin{figure}[t]
	\centering
	\includegraphics[width=17.2cm]{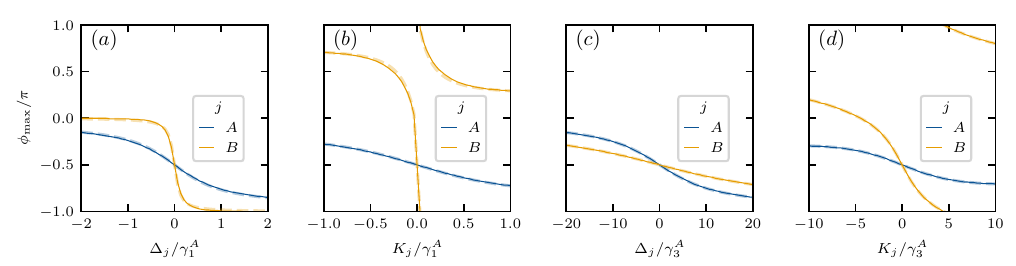}
	\scaption{Phase locking of standard limit cycles. Solid curves denote the maxima of $P_1$ whereas dashed curves represent $\text{arg}(\langle \tilde{a}_j\rangle)$. (a), (c) $K_j=0$. (b), (d) $\Delta_j=0$.
		In panels (a) and (b), we set the drive strength to $\Omega_j=\gamma^A_1/6$ and the gain and damping rates to $\gamma^A_1=\gamma^B_1$, $\gamma^A_2=2.5\gamma^A_1$, $\gamma^B_2=0.1\gamma^A_1$, and $\gamma^j_3=\gamma^j_4=0$.
		In panels (c) and (d), we set the drive strength to $\Omega_j=\gamma^A_3/6$ and the gain and damping rates to $\gamma^j_1=\gamma^j_2=0$, $\gamma^A_3=\gamma^B_3$, $\gamma^A_4=3.75\gamma^A_3$, and $\gamma^B_4=0.15\gamma^A_3$.}
	\label{fig:phimax_standard_LC}
\end{figure}

\ssection{Phase locking of single standard limit-cycle oscillators}
\ssubsection{Single Driven Standard Limit Cycle}
In this section, we compare the synchronization behavior of a driven TLC to the case of two individually driven standard limit cycles of different radii.
We set $\gamma^j_3=\gamma^j_4=0$ to obtain standard limit cycles.
The steady-state solution close to the stable radius of each system for $g=0$ and $\gamma^A_1=\gamma^B_1$ is given by
\begin{align}
r_j = \sqrt{\frac{\gamma^j_1}{2\gamma^j_2}} + \epsilon r_{j,1} ,~~j\in\lbrace A,B\rbrace\,,
\end{align}
where $\epsilon=\Omega_j/\gamma^j_1$ and $r_{j,1} = -\sin(\phi_j)$.
The equations of motion for the individual phases read
\begin{align}
\dot{\phi}_j =& -\Delta_j - K_j\frac{\gamma^j_1}{\gamma^j_2} - \Omega_j\sqrt{\frac{2\gamma^j_2}{\gamma^j_1}}\cos(\phi_j)  + 2\Omega_j K_j \sqrt{\frac{2}{\gamma^j_1\gamma^j_2}}\sin(\phi_j)\,. \label{eq:phiMFpert12}
\end{align}
Thus, we expect $\phi_j=-\pi/2$ for $\Delta_j=K_j=0$ as well as a shift towards $\phi_j=\pm\pi$ ($\phi_j=0$) for small $\Delta_j>0$ ($\Delta_j<0$) and for small $K_j>0$ ($K_j<0$), as shown in \cref{fig:phimax_standard_LC}.
The value of the locked phase displayed in \cref{fig:phimax_standard_LC} is qualitatively different from the one of the two limit cycles of a TLC presented in \cref{fig:phimax}.
The locking phase of the larger limit cycle $B$ reacts stronger to increasing $\Delta_B$ or $K_B$ than the smaller limit cycle $A$.

Another way to define a single quantum limit cycle, i.e., one ring in phase space, is to set $\gamma^j_1=\gamma^j_2=0$ while keeping $\gamma^j_3$ and $\gamma^j_4$.
For this choice, we obtain the following perturbed steady-state radii
\begin{align}
r_j = \sqrt{\frac{3\gamma^j_3}{4\gamma^j_4}} + \epsilon r_{j,1},~~j\in\lbrace A,B\rbrace\,,
\end{align}
where $\epsilon=\Omega_j/\gamma^j_3$ and
\begin{align}
r_{j,1} = -\frac{16}{27}\left(\frac{\gamma^j_4}{\gamma^j_3}\right)^2\sin(\phi_j)\,.
\end{align}
The equation of motion for the individual phase reads
\begin{align}
\dot{\phi}_j =& -\Delta_j - K_j\frac{3\gamma^j_3}{2\gamma^j_4} - \Omega_j\sqrt{\frac{4\gamma^j_4}{3\gamma^j_3}}\cos(\phi_j) + \frac{16}{9}\Omega_j K_j \sqrt{\frac{4(\gamma^j_4)^3}{3(\gamma^j_3)^5}}\sin(\phi_j)\,.
\end{align}
The qualitative locking behavior is similar to the case  $\gamma^j_3=\gamma^j_4=0$ given by \cref{eq:phiMFpert12}.
The simple analysis of the solution of $\phi_\mathrm{max}$ of \cref{eq:phiMF} at $\Delta=K=0$,
\begin{align}
\partial_\Delta \phi_\mathrm{max} = -\frac{r}{\Omega}\,,\label{eq:slopeDelta}\\
\partial_K \phi_\mathrm{max} = -\frac{2 r^3}{\Omega}\,,\label{eq:slopeK}
\end{align}
reveals that the dependence of the locking phase on both $\Delta$ and $K$ is stronger for larger radii in the classical system.
However, the locking of the quantum analogue shown in \cref{fig:phimax_standard_LC}(c) is different: the locking phase of the limit cycle with a smaller radius reacts stronger to detuning than the one with a larger radius.

\ssubsection{Two Coupled Standard Limit Cycles}
We now compare the synchronization behavior of two coupled TLCs to the case of two coupled standard limit cycles of different radii.
Setting $\gamma^j_3=\gamma^j_4=0$ to obtain standard limit cycles, the steady-state solution close to the stable radii of each system for $\Omega_A=\Omega_B=0$ and $\gamma^A_1=\gamma^B_1$ is given by
\begin{align}
r_j = \sqrt{\frac{\gamma^j_1}{2\gamma^j_2}} + \epsilon r_{j,1} + \epsilon^2 r_{j,2},~~j\in\lbrace A,B\rbrace\,,
\end{align}
where $\epsilon=g/\gamma_1$ and
\begin{align}
r_{A,1} =& \sqrt{\frac{\gamma^A_1}{2\gamma^B_2}}\sin(\phi_{BA})=-\sqrt{\frac{\gamma^A_2}{\gamma^B_2}}r_{B,1}\,,\\
r_{A,2} =& -\sqrt{\frac{\gamma^A_1}{2\gamma^A_2}}\frac{3\gamma^A_2+2\gamma^B_2}{2\gamma^B_2}\sin[2](\phi_{BA})\,,\\
r_{B,2} =& -\sqrt{\frac{\gamma^A_1}{2\gamma^B_2}}\frac{2\gamma^A_2+3\gamma^B_2}{2\gamma^A_2}\sin[2](\phi_{BA})\,.
\end{align}
The equation of motion of the relative phase reads
\begin{align}
\dot{\phi}_{BA} =& -\delta + K\left(\frac{\gamma^A_1}{\gamma^A_2}-\frac{\gamma^A_1}{\gamma^B_2}\right) + g\cos(\phi_{BA}) \left(\sqrt{\frac{\gamma^A_2}{\gamma^B_2}}-\sqrt{\frac{\gamma^B_2}{\gamma^A_2}}\right) + 4\frac{g K}{\sqrt{\gamma^A_2\gamma^B_2}}\sin(\phi_{BA}) \nonumber\\
&- \frac{g^2}{\gamma^A_1}\sin(2\phi_{BA})\left(1+\frac{\gamma^A_2}{2\gamma^B_2}+\frac{\gamma^B_2}{2\gamma^A_2}\right)\,.\label{eq:MFpertphi2LC}
\end{align}

For $\delta=K=0$, we expect that the relative phase locks to $\phi_{BA}\in\lbrace 0,\pi\rbrace$ for $\gamma^A_2 = \gamma^B_2$ and to $\phi_{BA}=\pi/2$ ($\phi_{BA}=-\pi/2$) for $\gamma^A_2 > \gamma^B_2$ ($\gamma^A_2 < \gamma^B_2$).
These three cases coincide with the mean-field predictions for TLCs \cref{eq:phiMFTLCxxSupp,eq:phiMFTLCxySupp}, where $(\mathrm{in,in})$ and $(\mathrm{out,out})$ correspond to $\gamma^A_2 = \gamma^B_2$, $(\mathrm{in,out})$ to $\gamma^A_2 > \gamma^B_2$, and $(\mathrm{out,in})$ to $\gamma^A_2 < \gamma^B_2$.

\begin{figure}[t]
	\centering
	\includegraphics[width=5.5cm]{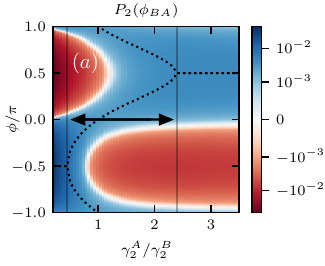}\hspace{1cm}
	\includegraphics[width=9.5cm]{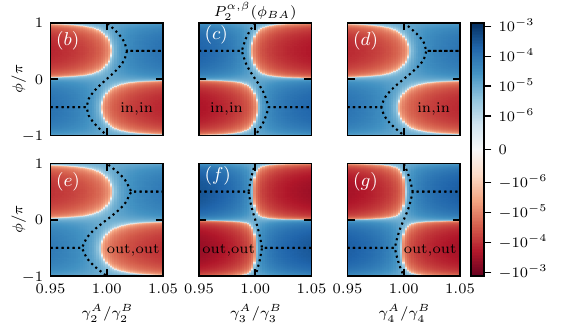}
	\scaption{Regions of the quantum synchronization blockade for $g=\gamma^A_1$, $\gamma^A_1=\gamma^B_1$, and $\gamma^B_2=2.5\gamma^A_1$.
		In all panels, maxima in $\phi_{BA}$ are denoted by dotted lines.
		(a) Two standard limit cycles with $\gamma^j_3=\gamma^j_4=0$ and $\gamma^B_2=2.5\gamma^A_1$.
		The blockade region is limited by the two vertical lines and indicated by the arrow (omitted in panels (b)--(g)).
		The color scale is linear in the interval $[-10^{-3},10^{-3}]$ and logarithmic elsewhere. (b)--(d) $P_2^{\mathrm{in},\mathrm{in}}$.
		(e)--(g) $P_2^{\mathrm{out},\mathrm{out}}$.
		In panels (b)--(g), the higher-order rates are set to $\gamma^B_3=1.04\gamma^A_1$ and $\gamma^B_4=0.096\gamma^A_1$ as in the main text.
		The color scale is linear in the interval $[-10^{-6},10^{-6}]$ and logarithmic elsewhere.}
	\label{fig:blockades}
\end{figure}

\ssection{Quantum Synchronization Blockade}
Here, we compare the stability of the synchronization blockade between coupled TLCs with the one of standard limit cycles, i.e., the range of dissipation rates $\gamma^A_j$ for which the blockade (bistable locking) persists.
The synchronization blockade arises for two identical oscillators (labeled $A$ and $B$) with equal gain and damping rates.
For two standard limit cycles, $\gamma^j_3=\gamma^j_4=0$, we fix both $\gamma^A_1=\gamma^B_1$ and $\gamma^B_2=2.5\gamma^A_1$.
Depending on the ratio $\gamma^A_2/\gamma^B_2$ of the damping rates, the synchronization measure $P_2(\phi_{BA})$ of the relative phase $\phi_{BA}=\phi_B - \phi_A$ defined in \cref{eq:P2viaOp} exhibits one or two maxima.
If at a given value of the ratio $\gamma^A_2/\gamma^B_2$ there are two maxima, the blockade (bistable locking) occurs, see the region between the vertical lines indicated by the arrow in \cref{fig:blockades}(a).
The location of the maxima of the synchronization measure is denoted by the dotted curve.
For $\gamma^A_2=\gamma^B_2$, the maxima are located at $\phi_{BA}=0,\pi$ as discussed below \cref{eq:MFpertphi2LC}.

When coupling a standard limit-cycle oscillator ($A$) to a TLC ($B$) and varying the ratio $\gamma^A_2/\gamma^B_2$ while keeping $\gamma^B_j$ fixed, no blockade emerges between the standard limit cycle and either of the limit cycles in the TLC, even if the mean-field values of the radii match.
Thus, the states need to be identical to observe the synchronization blockade effect between two oscillators.

In a system of two coupled TLC oscillators, we vary the rates $\gamma^A_2$, $\gamma^A_3$, and $\gamma^A_4$ individually by keeping $\gamma^A_1=\gamma^B_1$, $\gamma^B_2=2.5\gamma^A_1$, $\gamma^B_3=1.04\gamma^A_1$, and $\gamma^B_4=0.096\gamma^A_1$ fixed.
The resulting blockades are illustrated in \crefs{fig:blockades}(b) to \ref{fig:blockades}(g) and exist in a narrower range of $\gamma^A_j/\gamma^B_j\in[0.97,1.03]$ compared to the system of two standard limit cycles presented in \cref{fig:blockades}(a).
Thus, the blockade in a pair of TLCs is more susceptible to variations in gain and damping rates than that of two standard limit cycles.

\begin{figure}[t]
	\centering
	\includegraphics[width=17.2cm]{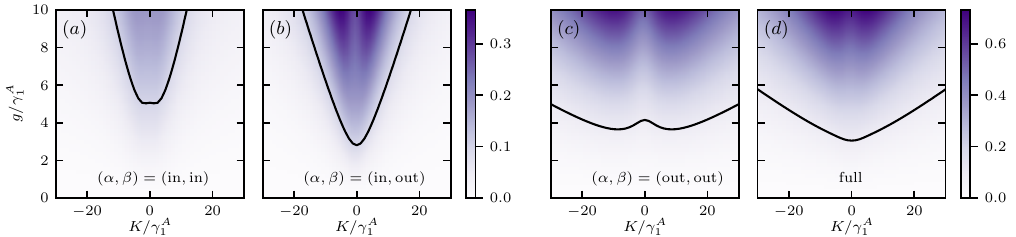}
	\scaption{Mutual information $I(\rho_{\alpha,\beta})\times 10^3$ for (a)--(c) truncated density matrices $\rho_{\alpha,\beta}$, see \cref{eq:rhotrunc}, and (d) the full density matrix $\rho$. The black curves denote contour lines at $0.1$.}
	\label{fig:mutual_info}
\end{figure}

\ssection{Quantum Mutual Information}
For a system consisting of two TLCs, we compare the quantum mutual information as a synchronization measure \cite{PhysRevA.91.012301} to $P_2^{\alpha,\beta}$ used in the main text.
The mutual information is defined as
\begin{align}
I(\rho) = S(\rho_A) + S(\rho_B) - S(\rho)\,, \label{eq:mutinfo}
\end{align}
where $S$ is the von Neumann entropy and $\rho_j$ are reduced density matrices.
For mixed states, the quantum mutual information contains both classical and quantum correlations.
To quantify the correlation between different limit cycles of the TLC oscillators, we truncate the density matrices as follows
\begin{align}
\rho_{\alpha,\beta} = \Tr{\mathcal{I}_A^\alpha\mathcal{I}_B^\beta\rho}\,, \label{eq:rhotrunc}
\end{align}
where $\mathcal{I}_j^\alpha$ refers to $\mathcal{I}^\alpha$ as defined in \cref{eq:IDdef} of the main text, which acts on the $j$th oscillator and the index $\alpha$ distinguishes between the inner and outer limit cycles.
In \cref{fig:mutual_info}, we show the mutual information for truncated density matrices as well as the full density matrix.
The behavior of mutual information is qualitatively similar to that of $P_2^{\alpha,\beta}$ shown in \cref{fig:arnold_tongues_g}.
It exhibits a dip around the blockade region at $K=0$ for the limit cycles (in,in) and (out,out), although it does not vanish completely. 
In contrast, the mutual information evaluated for the complete density matrix contains information about synchronization between different limit cycles and also the blockade effect between similar limit cycles.
Hence, the mutual information is not reduced significantly around the blockade.
Therefore, even if the mutual information reflects the blockade at $K=0$, it does not provide an advantage compared to the measures discussed in the main text.

\end{document}